\documentstyle[grlga]{article}
\lefthead{RIEDEL}
\righthead{STATISTICAL TESTS FOR
EARTHQUAKE PREDICTION}
\received{May 21, 1994}
\revised{September 15, 1994}
\accepted{October 31, 1995}
\journalid{GRL}{August 1994}
\articleid{1}{4}
\ccc{0148-0227/93/92JB-04601\$05.00}
\cpright{AGU}{1994}

\begin{document}

\title{Statistical Tests for Evaluating 
Earthquake Prediction Methods}

\author{Kurt S.\ Riedel}
\affil{Courant Institute of Mathematical Sciences \\ 
New York University}

\begin{abstract}
The impact of including postcursors in the null hypothesis test is
discussed.
Unequal prediction probabilities can be included in the null hypothesis test
using a generalization of the central limit theorem. A test for determining
the enhancement factor over random chance is given. The seismic earthquake
signal may preferentially precede earthquakes even if the VAN methodology
fails to forecast the earthquakes. We formulate a statistical test for
this possibility.

\end{abstract}
% Main Body Text

\section{Postcursors and High Significance Level}

%% -*- Mode: TeX; compile-command: ``latex grl''; 

The present paper
by {\it Varotsos et al.} [1996, hereafter cited as VEVL]
is devoted primarily to criticizing the criticism
of {\it Mulargia and Gasperini}
[1992, herafter cited as MG], although this paper does contain some
clarification of the  methodology
of {\it Varotsos et al.} [1981, hereafter cited as VAN].
The central argument is whether the VEVL method predicts better than random
chance. If the null hypothesis is true, the significance level is a
random variable with a mean value of $1/2$ 
and is uniformly distributed on [0,1].
Normally, one requires significance levels as low as 0.05 before one is
confident that the random chance hypothesis is false.

Many of {\it Mulargia and Gasperini's}  tests achieve
a significance level of .999, which means that under the MG test,
VAN do much worse than the hypothesis of random chance would expect.
There are three possible explanations for the  significance level of .999:
1) VAN are extremely unlucky; 2)  VAN predictions are anticorrrelated
with earthquakes; 3) the MG hypothesis that earthquakes are
independent and uniformly distributed in time is wrong.
Point 5 of the VEVL criticism suggests the likely culprit, i.e., many of MG's
earthquakes are probably aftershocks which VAN did not try to predict.
Note that the significance level of the test will converge to 1.0 as the
number of predictions tends to infinity if the null hypothesis of random
chance is true and a small percentage of the spatial distribution of
earthquakes are postcursors which VAN do not try to predict.
Thus the MG test should be repreated after excluding these postcursors. 
 It is unfortunate that VEVL did not
formulate a criterion for excluding postcursors.
% significance level, this suggests a
%systematic problem with MG's formulation. 

\newpage 

\noindent
\section{Spatial Distribution of Earthquakes}

MG do not include the limited spatial extent of the 
%\newpage \noindent
VAN predictions in
their analysis. Thus, MG are testing only the temporal accuracy and
not the spatial accuracy of the predictions. Such spatially averaged
tests cannot exclude the possibility that VAN can accurately predict
the location of the next earthquake, but has little success predicting
the time of occurence.
Nevertheless, testing the temporal accuracy is of interest.

The alternative,
including the spatial extent of the VAN predictions as
{\it Takayama} [1993] does, bears
the risk that
earthquakes are clustered on faults and that VAN are essentially predicting
occurences on the fault lines. By spatially
averaging, the MG analysis is
robustified against the spatial distribution of earthquakes. In contrast,
Takayama's analysis could depend quite sensitively on the spatial
distribution of earthquakes.

Since I am not a specialist, my only knowledge of the distribution of Greek
earthquakes is
from Fig.~19 of {\it Varotsos and Lazaridou}
[1991, hereafter cited as VL].
This figure makes the
earthquake distribution look highly nonhomogeneous, centered on the island
of Kefnalonia and running NW to SE. 

In estimating the spatial distribution of earthquakes, the entire
record of Greek earthquakes could be used i.e., earthquakes which 
occured before the VAN system was implemented could be used.
Including these earlier earthquakes assumes that the spatial distribution
of earthquakes has not changed appreciably over the time record.
In their response, I urge VEVL to plot both
the distribution of Greek earthquakes over the entire time record and
over the subrecord on which the VAN system was making predictions. 
It
would also be interesting to see the spatial distribution of VAN predictions
side by side.

From Fig. 19 of VL, a reasonable model for the spatial probability is
$$
p(\vec{x}) = p_0 + p_1
\exp ((\vec{x}- \vec{x}_c )^{\dag} Q(\vec{x}- \vec{x}_c )) \ ,
$$
where $\vec{x}_c$ is the most likely location of an earthquake, $Q$ is a
2$\times$2 positive definite matrix and $p_0$ is the spatially homogeneous
component of the distribution. Given the constraint that $\int p(x)dx = 1$,
this distribution has six free parameters. These parameters may be
estimated by maximizing the log-likelihood:
$\sum_{i=1}^N \log [p(\vec{x}_i )]$
where $\vec{x}_i$ is the location of the $i$-th earthquake.
Alternatively, the density can be estimated using bivariate kernel smoothers. 
(See {\em Silverman} [1986] and {\em Scott} [1992] for reviews.)

\noindent
\section{Null Hypothesis Testing with Unequal Probabilities}

Including the spatial inhomogeneity and the varying temporal lengths of the
predictions makes the probability of a correct prediction due to chance,
$j = 1 \ldots M$,
nonuniform. Let $p_j$ be the probability that the $j$-th forecast is
correct due to random chance. The nonuniformity makes the Poisson
summation formula nonapplicable. To evaluate the significance level
exactly is very computationally intensive. Monte-Carlo evaluations are also
costly. Instead, we recommend using Lindeberg's generalization of the
central limit theorem {\it Lamperti} [1966, p.\ 69]. 
In our context, it states that
$$
{( X_M - \sum_{j=1}^M p_j ) \over \sqrt{\sum^M_{j=1} p_j (1-p_j )} }
\stackrel{ M \rightarrow \infty}{\longrightarrow}  N(0,1) \ , 
$$
where the random variable, $X_M$, is the number of correct predictions
out of $M$ independent random attempts. We let $N_M$ be the number of 
correct predictions in the present data set.
Using these asymptotics,
the probability that $X_M$ is greater than or equal to $N_M$ is then
approximated by
$$
P(N_M \leq X_M ) =
$$
$$
P(N_M- {1 \over 2} \leq X_M ) \ \simeq \ 1 - \Phi
\left( {N_M- \mu_M - {1 \over 2} \over \sigma_M } \right) \ ,
$$
where
$$
\mu_M \equiv \sum_{j=1}^M p_j \ , \ \ \ \ \ \sigma_M^2 \equiv
\sum_{j=1}^M p_j (1-p_j ) \ ,
$$
and $\Phi$ is the cumulative error function. Since 
$\mu_M/\sigma_M$ increases as $\sqrt{M}$, 
this test will yield significance levels
 near zero as $M \rightarrow \infty$ even if the VAN method is only
slightly better than random chance. 

A more sophisticated question is: by what factor, $c$ is the VAN method
better than random chance? Here, $c=2$ means than the VAN success probability
is twice as large as random chance. The enhancement factor can be estimated
by 
%A crude statistic is
\begin{eqnarray*}
\hat{c} = N_M / \mu_M & =  & {\rm number \
predicted \ over \ number } \\
    &  & {\rm predicted \ by
\ random \ chance} \ .
\end{eqnarray*}

A more skeptical person would like to require a $2\sigma$ confidence interval
on the estimate of the enhancement factor. In other words, he/she would not 
accept the maximum likelihood estimate of $c$, but instead choose the
smallest value of $c$ that is consistent with the data.
Suppose the VAN
predictive probability were $p_j^{\rm (VAN)} = cp_j^{\rm (Random)}$, and
$\mu^{\rm (VAN)} = c \mu_M^{\rm (Random)}$ and $\sigma^{\rm (VAN)}_c =
\sum_{j=1}^M cp_j (1-cp_j )$. 
We assume that $ cp_j^{\rm (Random)}\le 1$ for each value of $p_j$.
We define the VAN predictive inflation factor as the minimal value of $c$
such that the {desired significance level (such as 0.05)} equals
$$
P_c (X_M \geq N_M- {1 \over 2}) = \ 1 - \Phi
\left( {N_M- c \mu_M - {1 \over 2} \over \sigma_c } \right) \ .
$$ 
When
$p_j < {1 \over 2}$, this probability inflation factor is less than the
simple ratio of $X_M / \mu^{\rm (Random)}$.

\noindent
\section{A Test for Precursor Activity}

It is possible that the seismic earthquake signals 
(SES) do tend to precede earthquakes, but that the VAN
algorithm does not forecast them well.
In particular, the temporal forecast might be inaccurate.
Thus we now change the question
from ``Is VAN better than random chance?'' to the more general question
``Do SES tend to precede earthquakes preferentially?''
Thus we test only whether the SES are earthquake precursors on
average and we do not test the temporal accuracy of the VAN forecast.

For simplicity, we neglect spatial dependencies and consider the time, $T$,
between the first prediction and the last earthquake. We assume that there
are $M$ predictions and $N$ earthquakes. We define $\tau_j$ to be the time
between the $j$-th prediction and the next earthquake.
Under the assumption that the first $N-1$ earthquakes are uniformly
distributed in time, $\tau_j$ is distributed as $p( \tau_j |t_j ) =
(N-1) (1- {\tau_j \over T} )^{N-2}/T$
for $0 \leq \tau_j$ $\leq$ $T - t_j$ and
$P( \tau_j = T-t_j ) = ({t_j \over T} )^{N-1}$, where $t_j$ is the time
of the $j$-th prediction. We use the test statistic $Y_M =
\sum_{j=1}^M \tau_j$.
The expectation of $Y_M = \sum_{j=1}^M E[ \tau_j ] = {T \over N}
\sum_{j=1}^M [1- ({t_j \over T})^N ]$.
The variance of $\tau_j$ is
\begin{eqnarray*}
Var [ \tau_j ] & = & T^2\left[{N-1 \over N^2 (N+1)} \right. \\
 & & \left. \,  + {2 \over (N+1)}
\left( {t_j \over T}  \right)^{N+1}
+{1 \over N^2}
\left(  {t_j \over T} \right)^{2N} \right]^2 \ ,
\end{eqnarray*}
but the $\tau_j$ are not independent. If we assume that $t_j$, $j=2,
\ldots , M$, are independently distributed in $[0,T]$ and average
over the distribution of $t_j$, the $\tau_j$ become independent.
In this case, we can apply the central limit theorem using
$Var [Y_M ] = \sum_{j=1}^M Var [ \tau_j | t_j ]$.
Let $\hat{\tau}_j$ be the observed delay time between the $j$ prediction
and the next earthquake.
If $\Sigma \hat{\tau}_j - E[Y_M ]$ is much smaller than $-$2.5
$\sqrt{Var [Y_M]}$, then the SES signals occur preferentially before
earthquakes. A similar test can be used to test whether the SES signals
tend to occur after earthquakes. We caution that this test will be sensitive
to the clustering of earthquakes and predictions into high activity time
periods.
If VEVL systematically exclude SES measurements which occur immediately
after an earthquake and the SES measurements are otherwise randomly
distributed, this test will confirm that the recorded SES signals occur
preferentially before an earthquake.

\noindent
\section{Risk Analysis and the VEVL Paradox}

This nonparametric test may be of value due to the ad hoc nature of the
VAN algorithms in predicting the spatial and temporal region where an
earthquake might occur. I strongly suspect that more detailed
statistical analysis could improve the forecasting algorithm given SES
measurements at various spatial locations.

The final topic which I would like to address is the paradox
in the appendix of VEVL.
The issue is: how does the uncertainty of predicting the
earthquake amplitude affect the hypothesis testing? VEVL's appendix raises
more general issues such as the relative loss of predicting an
earthquake which does not occur (or occurs at low amplitude) versus not
forecasting a large amplitude quake. A forecast is an action taken
under uncertainty, and an incorrect forecast will cause an economic cost.
To truly assess the effectiveness of an earthquake prediction scheme,
the loss associated with the scheme and the resulting policy decisions
needs to be compared with the costs of ignoring the prediction. When a
realistic assessment of the loss associated with predicting an earthquake
which does not occur is made, I believe that many prediction schemes
will be economically useless, but statistically significant. 

\acknowledgments
 The helpful comments of Dr.\ Gel\-ler and 
Dr.\ Vallianatos are acknowledged.
This work was supported by U.S. Department of
Energy Grant No.\ DE-FG02-86ER53223.

%  \input{grl3}

%% -*- Mode: TeX; compile-command: ``latex grl; xdvi grl''; -*-

%  \input{grl3}

\authoraddr{K.\ S.\ Riedel, 
Courant Institute of Mathematical Sciences
New York University \\
New York, NY 10012-1185. (e-mail: riedel@cims.nyu.edu)}

\end{document}